\documentstyle[aaspp4]{article}
\def\bm#1{\hbox{\boldmath $#1$}}

\input{psfig}

\begin{document}
 
%\twocolumn[$\;$\vspace{0.5cm}
\begin{center}\uppercase{{\sc Gas and Dark Matter Spherical Dynamics}}
\end{center}

\author{\sc Jean-Pierre   Chièze, Romain    Teyssier}
\affil{CEA, DSM/DAPNIA/Service d'Astrophysique, CE-Saclay,
       F-91191 Gif-sur-Yvette, Cedex, France}

\author{\sc Jean-Michel Alimi}
\affil{Laboratoire d'Astrophysique Extragalactique et de Cosmologie,
CNRS URA 173, Observatoire de Paris-Meudon, 92195 Meudon, France}

\begin{abstract}
  We  investigate the formation  of spherical  cosmological structures
  following  both  dark matter and gas   components.  We focus  on the
  dynamical aspect of  the collapse assuming  an adiabatic,  $\gamma =
  5/3$, fully ionized  primordial plasma.  We  use for that purpose  a
  fully  Lagrangian hydrodynamical  code  designed to  describe highly
  compressible  flows  in spherical geometry.   We  investigate also a
  ``fluid  approach'' to describe the  mean physical quantities of the
  dark matter flow.  We test its validity for a  wide range of initial
  density contrast.  We show that an homogeneous isentropic core forms
  in the gas  distribution,  surrounded by a  self-similar hydrostatic
  halo, with  much higher entropy  generated by shock dissipation.  We
  derive analytical expressions  for the size, density and temperature
  of the core,  as well as  for the surrounding  halo.   We show that,
  unless very efficient heating  processes occur in  the intergalactic
  medium,  we  are unable  to  reproduce within  adiabatic  models the
  typical  core sizes  in  X-ray  clusters.   We  also  show that, for
  dynamical  reasons only,  the   gas distribution is  naturally  {\it
  antibiased}  relative to    the  total mass    distribution, without
  invoking  any reheating processes.  This  could  explain why the gas
  fraction increases  with radius in very large  X-ray clusters.  As a
  preparation for the next study devoted to the thermodynamical aspect
  of the collapse, we investigate  the initial entropy level  required
  to solve the core problem in X-ray clusters.
\end{abstract}
%]
 
\section{INTRODUCTION}
\label{introduction}

This paper   is the first in   a series devoted to  the hydrodynamical
study of the formation of structures in cosmology.  We focus here on a
precise description of  the dynamics of  dark matter and gas collapse.
Non adiabatic  processes,  namely  temperature decoupling  between the
various plasma components, heat conduction, non equilibrium ionization
balance,    cooling,  heating by  star    formation are  examined in a
following paper (\cite{Paper2}).

Gas and dark matter distributions observed  in X-ray clusters provides
strong constraints  on  structure formation theory.   Although the gas
density profile shows in the most massive clusters a typical core--halo
structure (Hughes  1989;  Briel, Henry  \& Bohringer 1992),  the total
mass distribution deduced  from lensing analysis  seems to be strongly
peaked towards the  center, with no clear  evidence of a central core.
EINSTEIN and ROSAT observations of X-ray clusters  show gas core radii
with sizes  ranging from 100  to 300 h$^{-1}$  kpc.  Recent studies of
gravitational  lensing in  clusters   suggest  that the  total    mass
distribution can  be  singular in the center  (Wu  \& Hammer  1993 and
reference therein)  or at least  with very small core radii ($r_{core}
\simeq$ 20-30 h$^{-1}$ kpc)  (Tyson, Valdez \&  Wenk 1990;  Mellier et
al.  1994;  Flores \& Primack   1996).  Moreover, gas and dark  matter
distribution   differ  also in   the outer  parts of  galaxy clusters.
David, Jones \&  Forman (1995) (hereafter DJF)  show  that there is  a
strong {\it antibias} of the X-ray gas in the clusters with respect to
the  underlying total mass  distributions.    The  gas mass   fraction
increases   with radius,  up  to what  should be    the primordial gas
fraction $\Omega_B / \Omega_0 \simeq 0.2 -  0.3$ (White et al.  1993).
The origin of  such  an antibias is   an open  question.  A  model  of
cluster formation, which takes into account the  dynamics of both dark
and gas component, should be able to reproduces these features.

The most rigorous approach  so far of  the coupled gas and dark matter
dynamics  was performed  by  Bertschinger (1985)  in  the case of  the
self-similar secondary infall.  Its analysis showed that the final gas
distribution traces  the dark matter  distribution.  For  more general
initial conditions, it  is  generally admitted that such  a conclusion
subsists as  long as thermal  processes are not included.  More recent
studies, based essentially on 3D   numerical simulations of  structure
formation,  show however that  gas  and  dark  matter have   different
distributions    due to   the  hierarchical    nature of gravitational
clustering.  Antibiasing is due to a  systematic energy transfer from
dark matter  to gas during mergers of  protoclusters (Navarro \& White
1993; Pearce, Thomas \& Couchman  1994).  We propose to address  again
this  question, by considering first  the role of the purely adiabatic
dynamics  (this  paper), and    second the  role of   various  thermal
processes   (\cite{Paper2})  with   a  high resolution  hydrodynamical
treatment.

The  hierarchical    nature of  gravitational   collapse  requires, in
principle, high resolution,  3D  simulations.  But it is  difficult to
reconcile  physical   completeness  with topological  realism.   Plain
hydrodynamics simulations generally suffer from limited resolution and
reduced physical inputs,  while  one-dimensional calculations offer  a
practically unlimited  resolution, with  however a highly  constrained
and unrealistic geometry.  They are also useful  tools to work out and
reveal the potential effects of various physical processes.  Moreover,
accurate numerical methods  can be more easily  refined  in 1D models,
especially   in a  high  resolution Lagrangian    framework,  and then
provides  a valuable complement to more   elaborated 3D simulations of
the formation of large--scale structures.

The object of this paper is to study the gas and dark matter dynamics,
with  a one dimensional, high resolution,  Lagrangian  model, in a way
similar to  Thoul  \& Weinberg   (1995)  in spherical geometry   or to
Shapiro \&  Struck-Marcell (1984) in slab  geometry.  In the spherical
case presented in this paper, we focus  on the differences between gas
and   dark matter distributions due  to  dynamical processes only.  In
order to understand clearly the dynamical features of both components,
it appears more convenient to consider separately the pure gas and the
pure dark matter collapse, before investigating the coupled dynamics.

When the calculation of the pure gas collapse of a perturbed spherical
system is worked  out along these lines, it  appears that the collapse
of the inner regions actually  proceeds to very  high densities.  This
is due to the fact that -  given any reasonable density perturbation -
the inner subsonic  flow  is not far  from  homologous, so  that shock
heating has here little effects.  More precisely,  it exists a central
core  which   evolves   strictly adiabatically   up   to   hydrostatic
equilibrium.  This core is bounded by the sonic  radius (where the gas
velocity is equal to  the  sound velocity),  from which the  accretion
shock  originates and  propagates outward.  The  final central density
depends on  the core initial  entropy.   We develop in  Section 2.2 an
analytical model which describes  precisely the core structure and the
surrounding  halo.   Section   2.3  presents    results  of  numerical
simulations  which  confirm both  the   numerical  model developed  in
section 2.1 and the previous analytical model of a pure gas collapse.

The collapse of  collisionless dark matter is  supposed to provide the
potential well in   which gas is falling,   and thus deserves  special
attention.   The very high  dynamical range required for the treatment
of the  gas component claims for an  equivalent dynamical range in the
treatment of the collisionless  component. A classical method, adapted
to spherical geometry, consists   in calculating the  trajectories  of
concentric  shells of  collisionless   material.  However, it  appears
rapidly  that the numerical accuracy  which  can be achieved with this
method  cannot, for any   reasonable  computing time,  meet the   very
stringent  dynamical range and  resolution requirements imposed by the
gas evolution.  First, the dark matter distribution in each Lagrangian
gas element must be well sampled by a sufficient number of dark matter
shells, which   requires   a very  large  number  of  shells, with  no
guarantee  of a  smooth dark matter  distribution.    Second, the mass
distribution must be calculated over more than $10$ decades in radius.
In view  of  these difficulties,  we propose  in Section  3  a ``fluid
model'' for the dark matter component,  which very accurately accounts
for its gravitational behavior.    In this  way, the  collisional  and
collisionless components are treated on an equal numerical footing.

The resulting coupled model contains all the  dynamical aspects of the
spherical collapse of both gas and dark matter.  Some salient features
of this model  are discussed in  Section 4.  We  show  that for purely
dynamical   reasons   the gas  and   dark  matter distributions differ
significantly,  leading   to  an   {\it  antibiased} gas  distribution
relative to   the  total  gravitational   mass.   We  investigate  the
influence of the baryons density parameter on the gas distribution and
we compare  our results with  the  observational data of  DJF, finding
excellent  agreements  between our  model  and the  most massive X-ray
clusters.  The difficulty to form  realistic cores, even in  adiabatic
conditions, are  finally discussed, in  order to evaluate  the initial
entropy level required to explain the observed sizes of X-ray clusters
core radii.  It  turned out that  photo--ionization  processes are not
sufficient, and  that galactic winds  are good  candidates to transfer
enough energy to the intergalactic medium and solve  the problem.  The
more  complex and richer   thermodynamical aspects of cosmic structure
formation are discussed in a forthcoming paper (\cite{Paper2}).

\section{GAS DYNAMICS}

In  this section, we analyze  the basic properties  of the dissipative
heating of the   gas  during gravitational  collapse.   The  adiabatic
dynamics does not depend on the  ionization degree of matter.  The gas
is  here assumed to be  fully  ionized, composed  of 76\% hydrogen and
24\% helium by mass.  We consider here the gravitational collapse of a
purely gaseous, gravitationally bound, sphere. Emphasis  is put on the
treatment of shock   waves,   especially on the   conditions  of their
appearance, both in time  and  space.  Our high resolution  Lagrangian
treatment puts  forward the formation  of an adiabatic -  isentropic -
high density  core, in which  shocks never occurred. The importance of
this core is that it determines the rest of the structure, essentially
heated up  by the  dissipation of supersonic   infall motions,  in the
relaxation layer of a moving accretion shock.

Mass motion  kinetic energy is dissipated  into heat across this shock
by  the action of viscous forces.   The dissipative shock region has a
characteristic width of a few particle mean free paths, generally well
beyond the   resolution   of an hydrodynamical   simulation.   In most
treatments,  shock    transitions  are   handled   with   a  so-called
``artificial viscosity''  by  which the   entropy  production rate  is
largely increased  above its actual value,  so that the shock width is
spread over few  cells.  The success  of the  operation relies on  the
fact   that  Hugoniot-Rankine relations,   which  connect upstream and
downstream   gas  properties   (density,   momentum  and   energy) are
conservation relations, independent of the dissipative processes which
drive the transition.  However, it is essential to respect the correct
dependency  of the  viscous force  on   the structure of the  velocity
field, in   order to determine  safely where   and when in   the flow
``artificial''   viscosity  is   required.   The  viscous  forces  are
described by the stress tensor,  which, in spherical geometry, depends
only on the combination $\left({\partial u\over \partial r}-{u\over r}
\right )$.   As required, the   stress vanishes for  locally isotropic
flows.   This  is of great  importance  for  cosmological flows, which
spend a long time close to uniform expansion or contraction.

We present briefly the main features of the numerical code we employed
for this study. It is an evolution of  a code which has been developed
for various   astrophysical purposes  (\cite{Chieze2}; \cite{Chieze3};
\cite{Chieze1}).

In  order to   get  very high   resolution, the   numerical scheme  is
Lagrangian.  In all   calculations  presented  here, the  number    of
Lagrangian  shells is 250, both  for gas and   dark matter.  Since the
resolution    must  be concentrated  in    the core, shell  masses are
distributed according to  a geometrical progression, with  a dynamical
range of  $10^{10}$.   For example, in the   simulation of a $10^{16}$
M$_{\odot}$ cluster, the inner shell  has a mass of $10^4$ M$_{\odot}$
only, while the outer shells have a mass of $10^{14}$ M$_{\odot}$.  It
turns  out that increasing the number   of shells is unnecessary.  The
effective  spatial  resolution  in the centre  (in   our case 0.1  pc)
depends on the inner shell masses, not on  the total number of shells,
which just governs the resolution in the outer regions.

Numerical  stability is guaranteed by  a  fully implicit method, which
allows to retain the time step which limits the relative variations of
the  dependent variables to some  small arbitrary value, generally 1\%
in the simulations presented here.  The  independent variables are the
time $t$, and the cumulative mass $m$ from  the center.  The dynamical
dependent  variables are the  radius $r(m,t)$ of the sphere containing
the mass  $m$ and the corresponding  velocity  $u(m,t)$.  Dealing with
perfect gases of adiabatic  exponent $\gamma=5/3$, the thermodynamical
variable we adopt is the function

\begin{equation}
s(m,t) = \ln \left ( EV^{{1\over{\gamma-1}}}\right )
\end{equation} 

\noindent   
where $E$ is the  internal energy of  a fluid  element of volume  $V$.
The evolution  of  this thermodynamical  function,  reminiscent of the
entropy, is given by

\begin{equation}
{ds\over dt}= {\dot{\cal Q}\over E }
\end{equation}

\noindent 
where $\dot{\cal Q}$ is the variation  rate of the internal energy due
to dissipation or other thermal processes,  excluding the work done by
the  {\it thermal} pressure.   A choice of this  nature is dictated by
energy conservation considerations : it is clear that isentropic flows
$(\dot{\cal Q}=0)$  numerically remain  strictly isentropic, since  we
avoid the integration  of the work  done by  pressure forces over  the
whole history   of the system,   which invariably introduces numerical
inaccuracy.     The internal    boundary   conditions  are   naturally
$r~(0,t)=0$  and $u~(0,t)=0$ (no artificial  core  radius).  Since the
perturbation  we study  are always compensated,   the  velocity at the
outer boundary is taken to be  the unperturbed Hubble flow.  Our study
is  devoted  to   adiabatic  flows. Therefore, the    dissipation term
$\dot{\cal Q}$ is due to shock heating only.  We now describe the form
one should use for $\dot{\cal Q}$ in spherical geometry.

\subsection{Dissipation in Shock Waves} 

We  choose  to  handle shocks  by    the pseudo--viscosity  method (PV
method),  introduced for planar  shocks   by Von Neumann \&  Richtmyer
(1950).   Though it is often  considered  as a  numerical trick,  this
method  has in  fact firm physical   grounds.   In the  PV method, the
particle cross--section   is  reduced so  that  their  mean free--path
matches the cell size of the  simulation.  Consequently, the PV method
consists in solving correctly the  Navier--Stokes (NS) equations  with
the available resolution,  adapted in fact to  a highly  viscous flow.
Note that   the  Hugoniot--Rankine   relations,   which connect    the
downstream to  the   upstream  regions  of  an adiabatic  shock,   are
independent  of the detailed  microphysics,   since they just  express
conservation   laws.  This  is  the  reason   why the shock  adiabats,
calculated by the PV method,  are strictly identical to those obtained
by a NS  calculation (see Chièze,  Teyssier \& Alimi 1997  for further
comments).

In a fluid element, two distinct groups of time scales can be derived,
related respectively to the  microscopic properties of the plasma  and
to  the macroscopic properties  of the flow.   In  the first group, we
find  in particular  the mean collision   time $\tau  _{coll}$ between
particles, and  the  relaxation time  $\tau  _{rel}$ of  the  velocity
distribution due to collisions.  In  the second group, two time scales
characterize  the rate at which the  {\it shape}  of the fluid element
changes, and the rate at which its {\it volume} varies.  The ratios of
the time scales of  different groups, $\tau_{(1)}/\tau_{(2)}$, measure
the departure  from thermodynamical equilibrium,  which results in the
apparition of a viscous stress of magnitude $\Pi \simeq p ~ \tau_{(1)}
/ \tau_{(2)}$, where $p$ is the thermodynamical pressure.

Departures from thermodynamical  equilibrium resulting from the finite
relaxation time  of  internal degrees  of  freedom, give  rise to  the
so--called bulk viscosity.  The corresponding scalar correction to the
thermodynamical pressure is proportional to $\bm  \nabla \cdot \bm u$.
Since the relaxation   time of the energy  levels  of atoms  and  ions
($\simeq 10^{-6}  \hbox{s}$)  is negligible  compared to any  relevant
dynamical  time scale pertinent to the  problem, the bulk viscosity is
absent.  

The   usual viscous stress  tensor  is  related  to the  local rate of
strain, that is, the  rate at which the shape   of a fluid  element is
altered.  This means  that locally isotropic  motions can not generate
any viscous  stress.   In Cartesian coordinates,  the viscosity stress
tensor,  $\sigma   _{ij}$, and  the   dissipation rate   have the form
(\cite{Landau})

\begin{equation}
\sigma _{ij} =-\eta ~\left (
  {\partial u_i\over \partial x_j}+{\partial u_j\over \partial x_i}
 -{2\over 3}\delta_{ij} \bm \nabla \cdot \bm u \right ) 
\quad \hbox{and} \quad 
{\dot {\cal Q}}=\sigma_{ij}{\partial u_i\over \partial x_j}
\end{equation}

\noindent 
where  $\eta$ is   the viscosity   coefficient,  proportional  to  the
particles  mean free--path.      Since  we  are interested    here  in
spherically symmetric     flows, we  recast   these expressions   into
spherical coordinates

\begin{equation}
\sigma_{_{rr}}=2\eta~\left( 
          {\partial u\over \partial r} -{1\over 3}\bm \nabla \cdot \bm u 
                     \right)
\quad \hbox{and} \quad
\sigma_{_{\theta \theta}}=\sigma_{_{\phi \phi}}=
                           -{1\over 2}~\sigma_{_{rr}}
\label{radialstress}
\end{equation}

\begin{equation}
{\dot {\cal Q}}={3\over 2}\sigma_{_{rr}}
\left ( {\partial u\over \partial r}-{1\over 3}\bm \nabla  \cdot 
\bm u \right )
=3\eta \left ( {\partial u\over \partial r}-{1\over 3}\bm \nabla
\cdot \bm u \right )^2
\label{goodvisc}
\end{equation}

\noindent 
The equation of motion has the form

\begin{equation}
\rho {d u\over dt}=-{\partial P\over \partial r}
                   -\rho {\partial \Phi\over \partial r}
                   +{\partial \sigma_{_{rr}}\over \partial r}
                   +{3\over r}\sigma_{_{rr}}
\label{equmotion}
\end{equation}

\noindent 
where  $du/dt$ is   the  material acceleration  of    the fluid.  This
formulation  of  viscosity is of common   use in the   study of highly
non--homologous flows (e.g. Tsharnuter  \& Winkler 1979; Cioffi, McKee
\& Bertschinger 1988).

Compared to the above NS  solution, the single  manipulation in the PV
method concerns the viscosity coefficient $\eta$, which is used in the
simulation.  Since the   actual particle mean free--path $\bar{l}$  is
replaced by the local cell size, $\Delta x$, the viscosity coefficient
is increased by the factor $\Delta x / \bar{l}$.  Note that the radial
component of the   stress  (eq.  [\ref{radialstress}])  can be  either
positive or negative, which represents respectively a ``tension'' or a
``pressure''.  Since shocks lead always to compression, viscous forces
act as a pressure.  Artificial viscosity should then operate only when
the velocity field obeys the conditions

\begin{equation}
3 {\partial u\over \partial r} < \bm \nabla \cdot \bm u < 0
\end{equation}

\noindent
This completes the prescription that we  use to handle shock waves and
to avoid spurious viscous effects.

\subsection{Pure Gas Adiabatic Collapse}

Let   us      consider    first   an     adiabatic    gas     collapse
($\Omega_0=\Omega_B=1$). Note  that our results,  obtained  here in an
Einstein--de  Sitter universe,  are  also valid  for different models,
provided that the appropriate change of the cosmic  time are made.  We
show analytically and we confirm numerically that  a high density core
forms,  where  no  shock dissipation  has   occurred, surrounded  by a
self-similar hydrostatic halo, where shock dissipation is very strong.

\subsubsection{Initial Conditions}

We start at  an early epoch $t_i$,  into  the linear regime.   At that
time,  every spherical shell has  an initial  radius $r_i$.  We assume
that the  velocity field is unperturbed,  relative to  the homogeneous
expansion $v_i = H(t_i)r_i$, and that the initial density contrast has
the form

\begin{equation}
\rho _i = \bar \rho _i \left( 1 + \delta _0 \frac{ \sin q }{q} \right)
\end{equation}

\noindent
This density perturbation does not  correspond to a pure growing mode.
As long as the initial time is choosen early enough, the decaying mode
vanishes  quickly.  The Lagrangian  coordinate of each spherical shell
is defined  as $q = \Phi  _S r_i/R_S$ where $\tan  \Phi _S  = \Phi _S$
($\Phi _S \simeq 1.43 \pi$).  The  total mass of the  system ($M = 4/3
\pi \bar \rho _i R_S^3$) is then unperturbed.   The collapse epoch  of 
the system  is defined as  the first shell-crossing time, which writes
for these initial conditions

\begin{equation}
t_c = \frac{3\pi}{4} \frac{1+\delta _0}{\delta _0^{3/2}} t_i
\end{equation}

\noindent 
The two  independent parameters which   fix  then the  history of  the
system are the  total mass  $M$  and the collapse  epoch  $t_c$.   The
initial {\it uniform} temperature  of the gas  is given by the  cosmic
radiation  background,  which is tightly coupled   to  matter from the
recombination epoch at $z = 1000$ up to $z = 200$. This leads for $z_i
\le 200$ (see e.g. Anninos \& Norman 1994)

\begin{equation}
T_i = 546 \left( \frac{1+z_i}{201} \right)^2 ~ \hbox{K}
\end{equation}

\noindent
In Section 4, we allow  for much  higher  temperature due to  possible
reionization  processes by  stars  or quasars  UV background radiation
fields.

\subsubsection{Core Formation}

As long as the pressure  gradient term remains  small compared to  the
gravitational term (eq.  [\ref{equmotion}]), the pressureless solution
of spherical  collapse  can be used.  This   regime is valid  until $t
\simeq t_c$.  At that time, it is possible to derive {\it precollapse}
analytical solutions (\cite{Moutarde95}).   With the  notations we use
in this paper, this leads to the density profile

\begin{equation}
\rho (q,t_c) =  \frac{4}{7}\left( \frac{20}{3} \right)^2
 \bar \rho (t_c)  q^{-4}
\label{density}
\end{equation}

\noindent  
This behavior  is obtained for each initial  density profile for which
$\partial \rho_{i}  / \partial  r = 0$   and $\partial ^2  \rho_{i}  /
\partial r ^2 \ne  0$ at $r_i  = 0$.  Note  that  such a power  law is
valid for $q   \ll 1$ and  $t \simeq  t_c$.   A useful result   is the
position of the Lagrangian fluid element $q$ at $t \simeq t_c$.

\begin{equation}
r(q,t_c) = \left( \frac{9\pi}{5} \right)^{\frac{2}{3}}
\frac{R_S}{4 \Phi _S \delta _0}  q^{\frac{7}{3}}
\label{radius}
\end{equation}

\noindent 
One therefore deduces the  precollapse density profile  $\rho \propto
r^{-12/7}$.  During this pressureless infall,  each fluid element  is
adiabatically heated to a temperature given by

\begin{equation}
T(q,t) = T_i \left[ \frac{\rho (q,t)} 
{\bar \rho (t_i)} \right] ^{\frac{2}{3}}
\label{temp}
\end{equation}

\noindent 
which  allows  us to  compute  the local  sound   speed for each fluid
element $c_s(q,t_c) \propto  T(q,t_c)^{1/2}$. As the velocity field of
each infalling fluid element is

\begin{equation}
u(q,t_c) = - \left( \frac{9\pi}{5} \right)^{-\frac{1}{3}}
\frac{R_S}{4 \Phi _S \delta _0} \frac{\pi}{2t_c} 
q^{\frac{1}{3}}
\end{equation}

\noindent 
we can deduce the Mach number of each  fluid element defined as ${\cal
M}(q,t_c) = - u / c_s $.

\begin{equation}
{\cal M}(q,t_c) = \left( \frac{5\pi}{7} \right)^{-\frac{1}{3}}
\frac{R_S}{4 \Phi _S \delta _0}\frac{2\pi}
{t_c\bar c_s (t_c)} q^{\frac{5}{3}}
\label{mach}
\end{equation}

\noindent  
The fluid  element  for which ${\cal  M}(q,t_c) =  1$ reaches its {\it
sonic  radius} at   $t  \simeq t_c$.   Its coordinate   is  of primary
importance in our discussion.  Let us call the region enclosed by this
particular shell the {\it core} of the forming cluster.  For the fluid
elements with  $q < q_{core}$,  the flow is  always subsonic,  and no
shock  dissipation  occurs  in  this  region.  The  flow  leads to  an
hydrostatic configuration with  nearly  uniform density, pressure  and
temperature.

Outside the core, the flow is supersonic,  with higher and higher Mach
number for increasing radii.  When shocks are  strong enough, we enter
another regime  of the flow  leading  to a  self-similar structure  we
identify as the  {\it halo}.  Its  features are discussed in the  next
section.   Between  the core and  the halo,   there is an intermediate
region where shocks are weak  and heating  is  a mixture of  adiabatic
compression and   shock dissipation.   This   ensures the   transition
between the    adiabatic regime (core), and  the   strong shock regime
(halo).   Because  the core     is  in hydrostatic  equilibrium,   its
features remain remarkably stable up to $z = 0$.

It is possible to derive analytically  within this adiabatic model the
core density, temperature and  radius. Using equation (\ref{mach}), we
deduce the   value   of  $q_{core}$.  We then   obtain   with equation
(\ref{radius}) the core radius of the cluster

\begin{equation}
r_{core} \simeq 7 \times 10^{-4}~\hbox{kpc}
\left( \frac{1+z_c}{3}           \right)^{-\frac{17}{10}}
\left( \frac{M}{10^{16}~M_\odot} \right)^{-\frac{2}{15}}
\left( \frac{T_0}{0.014~\hbox{K}}\right)^{ \frac{7}{10}}
\end{equation}

\noindent 
$T_0$  is the average  temperature of  the  unperturbed background  at
collapse epoch, adiabatically  extrapolated until  today.  $M$ is  the
cluster  mass and $z_c$  the collapse  redshift.  Note  that  the core
radius is very sensitive to   the  temperature.  The core entropy   is
indeed equal to  the  mean background entropy.   Its  size is directly
related to  the  chosen value of  the   initial entropy.  Because  the
average baryon  density in a purely baryonic  universe is fixed by the
cosmological  parameters, this  means  that the  core radius is mainly
determined by the value of the temperature just prior to the collapse.
The final  density  inside  the core  can  be  computed using equation
(\ref{density})

\begin{equation}
n _{core} \simeq  7 \times 10^{7} \hbox{cm}^{-3}
\left( \frac{1+z_c}{3}           \right)^{ \frac{21}{5}}
\left( \frac{M}{10^{16}~M_\odot} \right)^{ \frac{4}{5}}
\left( \frac{T_0}{0.014~\hbox{K}}\right)^{-\frac{6}{5}}
\end{equation}

This core density  depends strongly on  the collapse epoch, because of
the $\left( 1+z \right)^3$  dependence of the mean background density.
The core temperature is simply derived using equation (\ref{temp})

\begin{equation}
T_{core} \simeq  8 \times 10^{6} \hbox{K}
\left( \frac{1+z_c}{3}           \right)^{\frac{14}{5}}
\left( \frac{M}{10^{16}~M_\odot} \right)^{\frac{8}{15}}
\left( \frac{T_0}{0.014~\hbox{K}}\right)^{\frac{1}{5}}
\end{equation}

These scaling laws agree  quite well with  the numerical resolution of
the hydrodynamics equations, as we will see below.

\subsubsection{Halo Formation}

As soon as  the core has  been formed, Lagrangian  fluid elements fall
towards the center with supersonic  velocity, leading to strong  shock
heating.  Let us suppose  that a given  fluid element is shocked at an
epoch given by its own  collapse time, in the  limit  $q \ll 1$,  this
gives

\begin{equation}
t_s(q) = t_c \left( 1 + \frac{3}{20}q^2 \right)
\end{equation}

\noindent 
from  which the Lagrangian coordinate of  the shock front, at any time
after the collapse, is obtained as

\begin{equation}
q_s(t) =2 \left(\frac{5}{3}\right) ^{1/2}
  \left( \frac{t}{t_c} - 1 \right) ^{1/2}
\end{equation}

\noindent 
Using equation (\ref{radius}), we obtain  the time dependence for  the
shock radius, which scales as $r_s \propto (t-t_c)^{7/6}$.  As soon as
the shock wave is strong enough (${\cal M} \gg  1$), it is possible to
estimate  the postshock temperature   $T_2$  and density $n_2$  using
Hugoniot-Rankine   relations.  If    we   assume that   the  preshock
temperature $T_1$ is negligible   (this is true  by definition   for a
strong  shock)   and that  the    postshock  velocity $u_2$  vanishes
(hydrostatic flow), this leads to the results

\begin{equation}
kT_2 = \frac{1}{3} \mu m_H u^2(q_s,t_c)
\quad \hbox{and} \quad 
n_2 = 4 \frac{\rho (q_s,t_c)}{\mu m_H} 
\end{equation}

After the shock front, because $u_2 (q)  = 0$, density and temperature
remain constant in time.  This implies  that the density profile after
shock  dissipation  has the same  slope  as  the  precollapse density
profile $\rho  \propto r^{-12/7}$ (in section  3, we will see that for
dark  matter the   picture  is  somewhat    different) and that    the
temperature  profile scales as $T  \propto r^{2/7}$, slowly increasing
with radius.  As the accretion shock escapes towards the outer part of
the  cluster, the  small $q$ approximation  breaks down  and  the halo
features are much more complex.

The   basic   approximation we make   here   is the  assumption  of an
hydrostatic postshock flow.  When we computed the core properties, we
have   also neglected  all   pressure effects  which  can  modify  the
trajectories.   These  points    have to  be tested   with   numerical
calculations which allow  us  to  solve precisely  the   hydrodynamics
equations.

\subsection{Numerical Calculations}

To test all the formulae presented in the upper  sections, and also to
illustrate  the importance of  the numerical prescription used in this
paper  to treat  shocks,  we run  a  simulation with the  same initial
conditions as previously stated. The   run parameters in this  section
are

\begin{equation}
M = 10^{16}~M_{\odot}~~~z_c = 2~~~T_0 = 0.014~\hbox{K}~~~\Omega_B = 1
\end{equation}

\begin{figure}[httb]
%\hspace{-1cm}
\hbox{\psfig{file=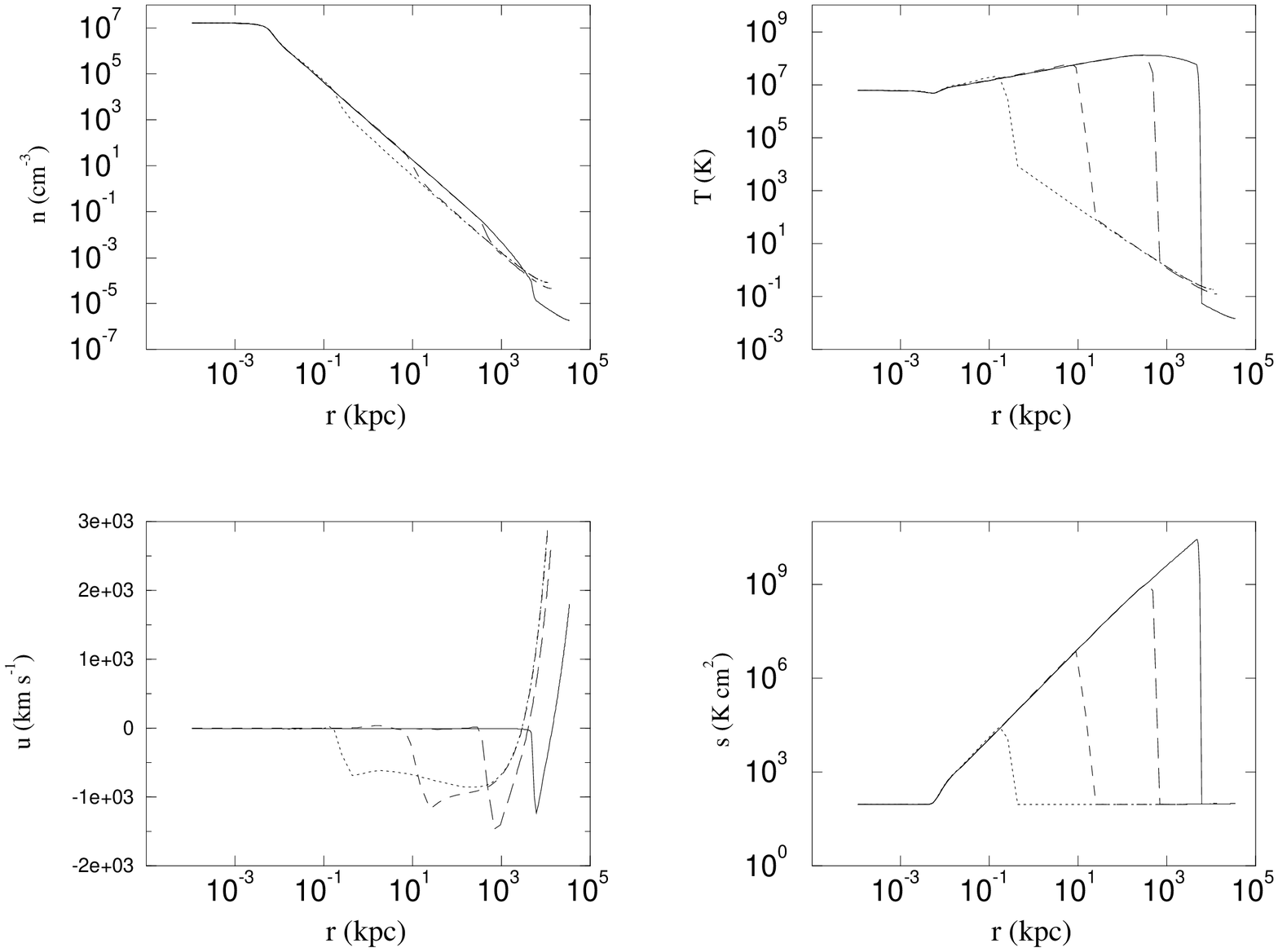,height=13cm,angle=270}}
\caption{Density, temperature, velocity and entropy profiles obtained
at  z=1.995 (dotted line), 1.98 (dashed  line), 1.5 (long-dashed line)
and 0  (solid line)  for the  pure  gas collapse with parameters  $M =
10^{16} ~M_{\odot}$, $z_c = 2$ and $T_0 = 0.014 \hbox{ K}$ .}
\label{figurebaryons}
\end{figure}

\noindent 
We    start   the simulation  at   $z_i=200$    and  reach $z=0$  with
approximately     $10^5$   time   steps.      We    plot  in    Figure
\ref{figurebaryons}    density, temperature,   velocity  and   entropy
profiles   obtained   at   different   redshifts,    using    equation
(\ref{goodvisc})  for the viscous  dissipation term.   The density and
temperature jumps at the shock front are just what is predicted by the
Hugoniot-Rankine relations.  The  assumption of  hydrostatic postshock
flow   appears to be valid  with  great  accuracy, since the postshock
velocity field is  zero  everywhere, and the density  and  temperature
profiles remains constant in time.  Note that no spurious oscillations
appears  in the post  shock  velocity field.   The power  laws for the
density  and  the temperature   profiles  in the  halo  are in perfect
agreement with  the analytical   derivation.   By the  virtue of   the
prescription for  viscosity (eq.   [\ref{goodvisc}]), the core remains
strictly   isentropic      during     the    collapse     (see    fig.
[\ref{figurebaryons}]).  Any other  prescription  would have led to  a
spurious core  heating, making impossible any  description of the most
inner part of the cluster.

More interesting is the good agreement  between the computed values of
the density and the temperature of  the core and the predicted values.
The  computed core density and temperature  are respectively $5 \times
10^{7}\hbox{cm}^{-3}$ and  $6   \times  10^{6}  \hbox{K}$, while   the
predicted  ones were $7   \times 10^{7} \hbox{cm}^{-3}$ and $8  \times
10^{6} \hbox{K}$. The computed core radius is about $2 \times 10^{-3}$
kpc while the predicted core radius  was $7 \times  10^{-4}$ kpc.  The
difference is larger for the   core radius, since pressure  gradients,
which   have been neglected  in   our  analytical approach, slow  down
infalling shells slightly before collapse. The important quantity here
is the  core  entropy,   determined   quite accurately by   both   the
analytical  and the numerical  calculations. Note that  the very small
size of the core  ($\sim 1$ pc) has  no  influence on the large  scale
properties of the flow.

When the entropy  profile starts to rise,   we enter the  halo regime,
where strong  shock  dissipation takes place.    When the  shock front
reaches outer regions of the cluster, the density profile get sharper,
since the small $q$  approximation is not  valid anymore.  The density
profile slopes from a $r^{-12/7}$ law in the  center to a $r^{-3}$ law
in the outer parts.

\section{A FLUID MODEL FOR DARK MATTER}

The  collisionless component of the universe  is  supposed to dominate
the total  gravitational potential  at nearly all  scales.   In a high
dynamical  range simulation, the  main difficulty  is  to get for dark
matter a resolution comparable to that achieved for the gas component,
ranging up to 10 orders of magnitude in density.

We   first approach the   problem   of dark   matter dynamics  by  the
traditional ``spherical  shells  model'' (Hoffman,  Olson \&  Salpeter
1980; Peebles 1982;  Hoffman, Salpeter   \& Wasserman 1983;   Hausman,
Olson \& Roth 1983; Martel \& Wasserman 1990; Thoul \& Weinberg 1995).
In this description, the  collisionless particles are distributed over
$N$ concentric shells of individual masses $m$ and radius $r$. As long
as the mass $M$ included below a given shell is constant, its dynamics
is determined  by Kepler's laws for  a  central potential motion.  The
highest     accuracy on  kinematics   and   energy conservation ($\sim
10^{-15}$ for a SUN station) is achieved  by using this exact solution
for each shell, up to shell crossing. At that time, new total energies
can be calculated, and new  exact trajectories are computed, until the
next  shell crossing time, and  so on.  This time--consuming procedure
guarantees  in  principle  a very   high    level of accuracy,   since
trajectories   are analytically treated.  However,   in the limit of a
continuous dark matter  distribution with zero angular momentum, shell
crossings    occur  continuously  at  the center    of  the  spherical
distribution, where the density   is  infinite.  With a finite   shell
number, the first  crossings (in radii)  occur in a  rapidly shrinking
region, the size of  which ultimately excludes any numerical solution.
Moreover, a correct  sampling of the  shell distribution in  the inner
regions (in particular over the size of the adiabatic core of the gas)
requires  an  excessive number of  shells, resulting   in a definitely
prohibitive computing time.

We have  also examined various  adaptations of the  Particle Mesh (PM)
technics, abandoning  the  quality of  the  semi-analytical  solution.
Again,  the very  small   inner scales  were not  well-sampled, mainly
because of the Eulerian nature of PM schemes.
  
We propose in  the following a fluid  treatment for dark matter, which
reproduces properly some aspects of the self-similar solutions of dark
matter  collapse, derived in    particular by Fillmore   and Goldreich
(1984) (hereafter FG) and  Bertschinger (1985). This unusual choice,
initially dictated by the  high resolution requirements  of Lagrangian
simulations,  provides  a new  insight   on collisionless dark  matter
dynamics.  Moreover, it prevents  for any two--body relaxation effect,
inherent  to any description  of dark matter with   a finite number of
shells.

\subsection{Dark Matter in Spherical Geometry}

The collisionless  dark  matter fluid is  described   by the velocity
distribution function $f(\bm r,  \bm v,t)$, which evolves according to
the Poisson--Vlasov  equation.  For a  spherical system  of  particles
moving along radial orbits, it reduces to $f(r,v_r,t)$, where $v_r$ is
the individual  particle radial  velocity.  The  hydrodynamical,  bulk
velocity of the fluid is $u =<v_r>$.

In a collapsing halo, three different regions can be distinguished:

\begin{itemize}
\item{The single stream, infall region, where the motion is purely 
Keplerian.  The distribution  function reduces to a $\delta$ function,
centered on the local flow velocity.}
\item{The relaxed core, where the dark matter fluid has been heated up
by phase mixing. We show in a companion paper (Teyssier, Chi{\`e}ze \&
Alimi 1997)  that, for the wide class  of  self--similar collapse, the
adiabatic invariance of  the gravitational  potential ensures in  this
region the evenness of the distribution function in velocity space. We
take advantage of this property to derive the equation of state of the
dark matter fluid.}
\item{The relaxation region, between the single stream and the relaxed 
regions, where   the velocity distribution can   only be  described by
solving the Poisson--Vlasov   equation.    We assume  here  that  this
dissipation region is of finite size, so that one can connect the flow
properties  on  both sides,     by  the  mass, momentum   and   energy
conservation laws. The resulting ``jump'' conditions are the analogous
of  the Hugoniot--Rankine relations   for classical fluids. Therefore,
the preceding discussion of dissipation  in fluids can be extended  to
dark matter.}
\end{itemize}

\subsection{Dark Matter Fluid Thermodynamics} 
\label{DMthermo}

We first derive the  equation of state  of dark matter, in the relaxed
region  where   our fluid  approach  is  valid, from  virial  and from
thermodynamical  considerations (see  also  Teyssier, Chièze \&  Alimi
1997 for further comments).

We consider a large system of particles, with an equilibrium spherical
distribution.   Particles are   moving   only along radial  paths,  as
required  by  the spherical   dark  matter  model.  Consider the   $N$
particles  interior to the  sphere  ($\Sigma$) of  radius $L$ and mass
$M$.    The aforementioned evenness   of  the distribution function is
equivalent  to assume a  quasi--stationary   state, with  no mass  and
energy flux through any surface element.  An equal number of particles
are  flowing in and out  the system, crossing ($\Sigma$) with opposite
velocities.  These two  opposite flows result in  a net momentum flux,
$p \bm n dS$, through any surface  element drawn on ($\Sigma$), with a
direction opposite to the normal $\bm  n$ to the surface element $dS$.
It is thus  without consequence to  materialize the surface ($\Sigma$)
by a reflecting sphere, on which the  exterior fluid exerts a uniform,
normal force   $-p \bm n$  per  unit surface.   Defining the net force
exerted on the surface ($\Sigma$) by the  fluid exterior to the system
by

\begin{equation}
-{\bar {\omega}}=-\int_{\Sigma}p~ \bm n\cdot \bm n~dS=-4\pi L^2p 
\end{equation}

\noindent  
one can see immediately  that the total kinetic  energy of the system,
$E$, obeys the virial equation

\begin{equation}
2E={\bar {\omega}}L
\label{eosdark}
\end{equation}

\noindent 
This  is the equation of  state of the  system, which has been derived
here from purely mechanical considerations,  assuming a even particles
distribution function.

The thermodynamical point of view leads to more specific results.  The
energy  $E$  and the   entropy $s$  are  the state   variables  of the
system. Both are extensive quantities relative  to the total number of
particles  $N$. Since each particle has  only one translational degree
of  freedom, the  entropy, function of  $N$,  $E$ and  of the external
parameter $L$, has the form
 
\begin{equation}
s~(E,L)=Nk~\left(\ln {L\over N}+{1\over 2} \ln {E\over N}\right)+s_0
\label{equentropy}
\end{equation}

\noindent 
The   first and   second  principle  of   thermodynamics  lead to  the
expression

\begin{equation}
dE = Tds - {\bar {\omega}}dL 
\end{equation}

\noindent 
where  one has defined the intensive  conjugated  variables of $S$ and
$L$ :

\begin{equation}
T = \left( {\partial E \over \partial s} \right)_L =2~{E\over Nk} 
\quad \hbox{and} \quad
{\bar {\omega}} = -\left( {\partial E\over \partial L} \right)_s
=2~{E\over L}
\end{equation} 

\noindent 
which is again the  equation of state of  the fluid.  Clearly,  for an
isentropic evolution, one has

\begin{equation}
\left ({\partial {\bar {\omega}}\over \partial \mu}\right )_s=3~
{{\bar {\omega}}\over \mu}
\end{equation}

\noindent 
where  we have  defined the   {\it  linear} mass density  $\mu  \equiv
\partial  M/   \partial L$.   It   can  be easily  shown   that $\left
({\partial  {\bar   {\omega}}/ \partial \mu}\right   )_S$  is just the
square of the sound velocity.  Specific heats are properly defined for
transformations respectively at constant ${\bar{\omega}}$ and constant
$L$.    From equations   (\ref{eosdark})  and  (\ref{equentropy})  one
readily  obtain the  corresponding specific heats $C_{{\bar{\omega}}}$
and $C_L$ as

\begin{equation}
C_{{\bar{\omega}}}={3\over 2}Nk ~~~~~ C_L={1\over 2}Nk
\end{equation}

With  this peculiar  understanding   of   the specific  heats,    this
monodimensional   gas can  be   considered as   a $\gamma^{(1)}\equiv
{C_{\bar{\omega}}/ C_L} = 3$ gas. The thermodynamical equilibrium of a
(gravitational) force free system  is characterized by the  uniformity
of $T$,  ${\bar  {\omega}}$ which  imply  a uniform {\it  linear} mass
density $\mu$.  In other words, the  equilibrium mass density ${\rho}$
and particle velocity dispersion $\sigma$ profiles are

\begin{equation}
\rho \propto L^{-2}  ~~~~~~  \sigma=Const.        
\end{equation}

\noindent 
However, starting from an arbitrary distribution of particles in phase
space, thermodynamical  equilibrium   can  only  be   reached  through
gravitational interactions,   since   particles are supposed    to  be
collisionless.

\subsection{Dark Matter Fluid Equation of Motion and Dissipation}

In the dark  matter spherical model,  where particles have no  angular
momentum, no   initial velocity dispersion,   and suffer no collision,
their  orbits of    zero eccentricity degenerate   into  purely radial
trajectories.   The   motion  of   an   individual  particle  is  thus
monodimensional,  contrary to a  classical  gas for  which collisions
populates the three  translational degrees  of freedom.  Defining  the
hydrodynamical   velocity $\bm u$     as the mean   local velocity  of
particles, and the   gravitational potential   as $\Phi  (r,t)$,   the
equation governing adiabatic fluid motions is immediately derived from
the definition of ${\bar{\omega}}$

\begin{equation}
{du\over  dt}=- \frac{1}{\mu}  {\partial {\bar{\omega}} \over \partial
             r} - {\partial \Phi \over \partial r}
\label{eulerdark}
\end{equation}

\noindent   
which express the conservation of the momentum of a spherical shell of
unit  mass.  The position  coordinate   is now  written  as usual   in
spherical  geometry     $r$.     The   equation     of   motion   (eq.
[\ref{eulerdark}]) is  hyperbolic and thus leads  to  the formation of
shocks.   It cannot   account for the  phase  mixing,  which  results,
according to  the Vlasov--Poisson approach,  from shell crossings.  As
already stated, we  assume that the  relaxation  region, identified as
the first leading caustics in the solutions of the Vlasov equation, is
of finite  size. The upstream  and  downstream flow properties are
connected by the  continuity of the mass,  momentum  and energy flows,
which write

\begin{eqnarray}
\mu_2 u_2 & = &
\mu_1 u_1 \\
{\bar{\omega}}_2 + \mu_2 u_2^2 & = &
{\bar{\omega}}_1 + \mu_1 u_1^2 \\
\left(\frac{3}{2}{\bar{\omega}}_2 + \frac{1}{2} \mu_2 u_2^2\right)u_2 & = &
\left(\frac{3}{2}{\bar{\omega}}_1 + \frac{1}{2} \mu_1 u_1^2\right)u_1 
\end{eqnarray}

\noindent
where the subscripts refer to the upstream and downstream regions.
Consequently, the ``jump''  conditions, which must be verified  across
the relaxation region, are

\begin{equation}
\mu_2 = 2 \mu_1 \quad u_2 = \frac{1}{2} u_1 \quad 
{\bar{\omega}}_2 = \frac{1}{2} \mu_1 u_1^2
\end{equation}

\noindent
Since they do  not depend on  the detailed  relaxation processes, they
are also the solutions, for our dark matter fluid, of the analogous of
the Navier--Stokes  equations (which however  do not provide the exact
solution {\it inside} the relaxation  layer).  Due to the symmetry  of
the system, the viscous  forces acting on   a fluid element  is purely
radial, so that the equivalent of the Navier--Stokes equation is

\begin{equation}
{du\over dt}=-{\partial\over\partial M}({\bar{\omega}}
             +{\bar{\bar{\omega}}}) 
             -{\partial \Phi \over \partial r}
\end{equation}
   
\noindent 
where   ${\bar{\bar{\omega}}}$ is the  viscous   force defined on  the
surface of the sphere of  radius  $r$. We  determine  now the rate  at
which energy is  dissipated  in a fluid   element  of unit mass.   The
variation of its kinetic energy, $K$, can be written as

\begin{equation}
{dK\over dt}= - {\partial({{\bar{     \omega}}} u)\over \partial M}
              - {\partial({ \bar{\bar{\omega}}} u)\over \partial M}
              - u{\partial \Phi \over \partial r}
              + {\bar{     \omega} }{\partial u\over \partial M}
              + {\bar{\bar{\omega}}}{\partial u\over \partial M} 
\end{equation}

\noindent 
The first, second  and third terms of  the rhs represent  respectively
the rates at  which the pressure,  viscous and gravitational forces do
work on the fluid element. This total amount of energy is not entirely
restored as flow kinetic energy. Indeed, the  two last terms represent
the energy which goes into heat, the first of which is the compression
work. Therefore, the expression

\begin{equation}
{\dot q}=-{\bar{\bar{\omega}}}{\partial u\over \partial M}
\end{equation}

\noindent   
represents the rate of increase of  the internal energy resulting from
the dissipation of the kinetic energy by viscosity.  Since dissipation
can only result in an entropy increase, one has furthermore

\begin{equation}
{\bar{\bar{\omega}}} {\partial u\over \partial M}<0
\label{irreversible}
\end{equation}

\noindent 
The expression of ${\bar{\bar{\omega}}}$ must be definite and positive
in dissipative regions,  that is in shell  crossing regions, near  the
leading  caustic,  and  zero   elsewhere.  In    this monodimensional
(radial) fluid, there is no distinction between the rate of strain and
the rate of  increase of  the length of   a linear element  of  fluid,
$\partial u/ \partial r$. For these reasons, the dissipative force can
be written as

\begin{equation}
{\bar{\bar{\omega}}}=-{\bar{\eta}}~{\partial u\over \partial r}
\end{equation}

\noindent 
which automatically  verifies the irreversibility  condition expressed
by equation (\ref{irreversible}), provided that ${\bar{\eta}}>0$. This
expression  of the  viscous force would  be in  principle valid for  a
collisional monodimensional  fluid,  for which  the particle velocity
distribution   is   governed by  collisions.  This    is not  true for
collisionless  material, where   the  particle  velocity  distribution
remains  completely  determined by  the initial  conditions  until the
occurrence of the  first   shell  crossing.   Until that stage,    the
collisionless fluid conserves   the memory of the  initial  velocity
distribution, contrary to a  collisional gas for  which this memory is
abolished after a few collision times. In our fluid description, it is
reasonable  to  consider that shell   crossing, and thus phase mixing,
actually occurs when the criterion

\begin{equation}
\frac{\partial u}{\partial r} - \frac{u}{r} < 0 
\end{equation}

\noindent 
is  fulfilled.  We shall see  in the following section that, according
to this  criterion, the position  of the fluid relaxation  layer (that
is, the  shock position) coincides quite  accurately with the position
of the leading caustic of collisionless material.

\subsection{Self Similarity Solutions for Dark Matter}

The role which  is essentially assigned to dark  matter  is to provide
the gravitational potential  well  in which gas eventually  condenses.
Therefore, two  main   requirements  must  be fulfilled  by    a fluid
treatment.  At  first, the over--all dark matter  mass  profile must be
correctly described. Second, the time dependence of  the growth of the
relaxation region must be also correct. We check here these two points
by  comparing our fluid model    results  to some known   self--similar
solutions of the problem of dark matter collapse.

Similarity solutions  have been found by FG  for the problem  of the
collapse  of  cold, collisionless  matter  in  a perturbed Einstein--de
Sitter universe. The   scale--free initial mass  perturbation they have
considered is of the form

\begin{equation}
{\delta M\over M} = \left ( {M\over M_0}\right )^{-\epsilon} 
\end{equation}

\noindent 
where $M_0$ is  a  scaling mass  and $0<\epsilon\leq  1$.  The initial
velocity field they have  considered  is unperturbed relative to   the
Hubble flow.   After shell crossing,  the dark matter density profiles
they obtained are well fitted by a power law $\rho(r)\propto r^{n}$ at
small radii, with, for spherical geometry, the following values for n

\begin{equation}
n=-2~\hbox{for}~~\epsilon \leq {2\over 3} ~~~\hbox{and}~~~
n=-{9\epsilon \over {3\epsilon + 1}}~\hbox{for}~\epsilon \geq {2\over 3}
\label{slopefillmore}
\end{equation}

\begin{figure}[httb]
%\hspace{-1cm}
\hbox{\psfig{file=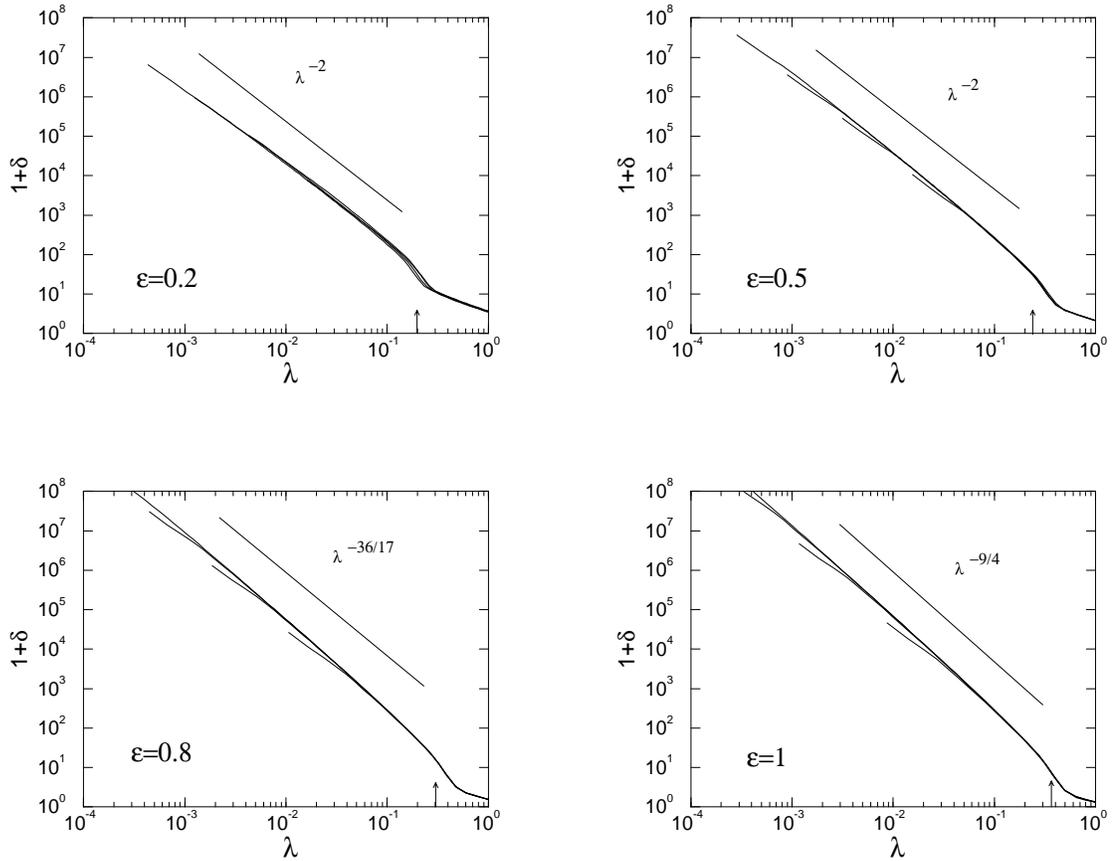,height=13cm,angle=270}}
\caption{Over-density profiles versus the self--similar radius $\lambda$ 
(see Section 3.4) obtained
at different redshifts for the pure dark matter case. We used scale
invariant initial conditions with $\epsilon$ = 0.2, 0.5, 0.8 and 1. The self
similarity is perfectly recovered, as well as the correct power law.
The position of the dark matter ``shock front'' (arrows) is exactly at the
radius of the first caustic calculated by Fillmore \& Goldreich (1984).}
\label{figautosim}
\end{figure}

We present  in  Figure  (\ref{figautosim}) the   overdensity profiles
obtained  at different redshifts for the  cases $\epsilon$ = 0.2, 0.5,
0.8  and 1.  The   radius have been   rescaled  at each  epoch to  the
self--similar radius defined as  $\lambda = r/r_{ta}$ where $r_{ta}$ is
the current turn--around  radius.  Except  naturally the caustics,  the
overdensity profiles we obtain  are exactly those presented in  FG.
In our  fluid model, dissipation  occurs for dark matter through shock
waves. The position of the shock front marks the first shell crossing,
and therefore has  to be close  to the first   caustic. We define  the
shock front  as the point where  the second derivative of the velocity
field reaches its minimum. We plot the position of this shock front in
Figure (\ref{figautosim}) as an arrow on the $\lambda$ axis. Note that
both the power law and the  position of the  first caustic are in very
good agreement between   our  fluid   model  and  the    collisionless
self--similar solutions.  Note also   that the dynamical range in radius   we
obtained is much  larger than  in  any other  dark matter simulations,
allowing us to achieve very high resolutions.

The case $\epsilon  = 1$ corresponds to the  secondary infall onto  an
initially overdense  perturbation embedded  in an Einstein--de  Sitter
universe.  Bertschinger (1985) has examined this situation in details,
both    for  collisionless and collisional     fluids.  Our result  is
quantitatively similar  to   the  relevant collisionless  overdensity
profiles presented  in Table  4 of  Bertschinger (1985).  This  author
found that  the position  of the  leading caustic  for dark matter  is
$\lambda = 0.364$. We recover this  value with 2\% accuracy within our
fluid model.  Bertschinger (1985) found  that the shock front position
for a $\gamma = 5/3$ gas was $\lambda = 0.339$, very close to the dark
matter  leading caustic.  He   noticed   however that $\gamma   =  3$,
pertinent  to a   one dimensional  gas,  should  have been  preferred,
although  he obtained in  that case  $\lambda  = 0.600$. This apparent
paradox has been elucidated here (section \ref{DMthermo}).

\subsection{Arbitrary Initial Conditions}

\begin{figure}[httb]
%\hspace{-1cm}
\hbox{\psfig{file=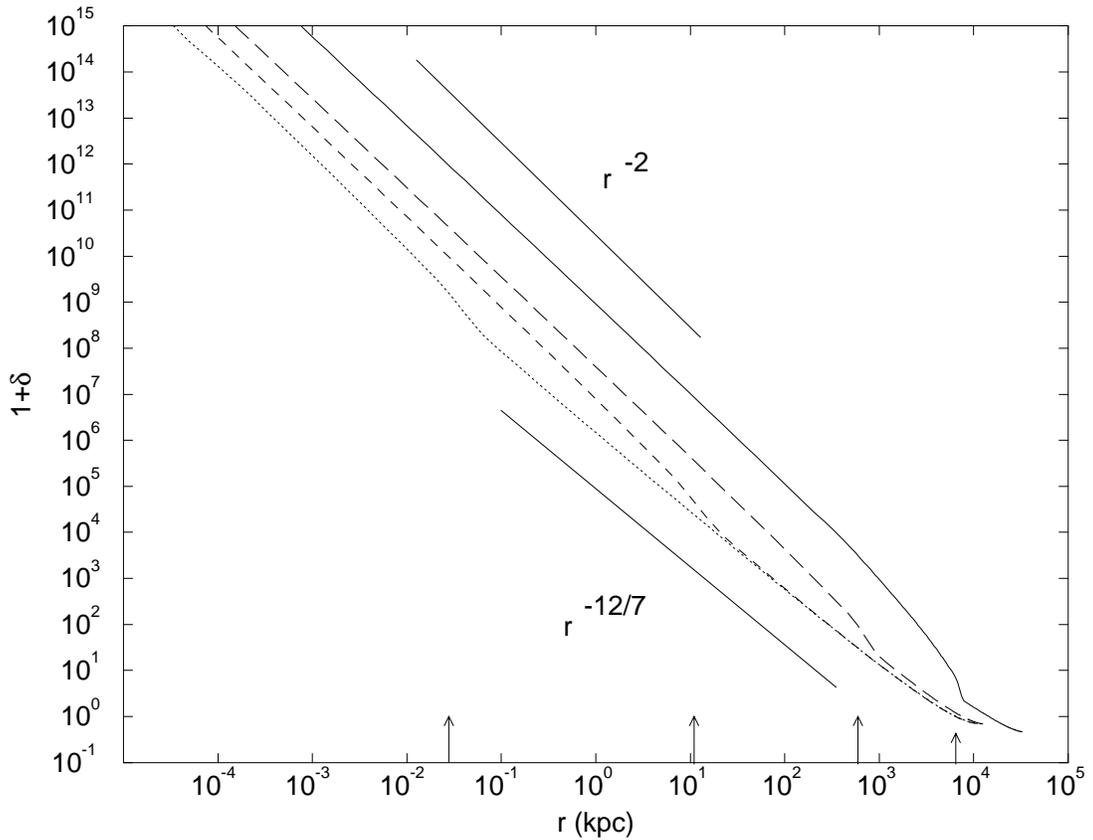,height=13cm,angle=270}}
\caption{Over-density profiles versus radius obtained at redshifts $z$ = 
1.995 (dotted line), 1.98 (dashed line), 1.5 (long-dashed line) and 
0 (solid line) for the pure dark matter case. We used a single Fourier
spherical mode as initial perturbation. Before the dark matter 
``shock front''(arrows), the pre-collapse slope $n=-12/7$ is recovered. 
After that, the corresponding self similarity solution of Fillmore \& 
Goldreich (1984) is obtained with a slope close to $n=-2$.}
\label{figpurecdm}
\end{figure}

The  similarity solutions for dark  matter collapse are well recovered
within    our fluid   model.   It  reproduces   also  the precollapse
scale invariance obtained  by Moutarde et al. (1995)  for a wide range
of initial   density  profile.   This precollapse  scale invariance
allows to built   up   naturally  the  conditions for    self--similar
collapse, for which our  fluid model is  fully valid.  As a matter  of
fact, we plot  in   Figure (\ref{figpurecdm})  the evolution  of   the
density profile  for a  pure  dark  matter  collapse, using  the  same
initial conditions as in  section  2.  Before  the first caustic,  the
density profile  follows the   precollapse  scale  invariance  $\rho
\propto r^{-12/7}$,   and after shell  crossings,  the density is well
fitted by the power  law $\rho \propto  r^{-2}$.   It is important  to
notice that the  mass distribution is not the  same than  for the pure
gas collapse.  Moreover, the postshock  velocity vanishes exactly  for
the gas, leading to a constant central density, while for dark matter,
the central density   increases continuously with  time, following the
time scaling  law given by  FG for  the case  $\epsilon = 4/9$,  which
corresponds to a  precollapse   density profile  with  a  slope   $n =
12/7$. At late epoch, and for large radii, the self--similar regime is
no longer valid, and a fully numerical treatment is necessary.

\section{COUPLED DARK MATTER AND GAS EVOLUTION}

In section  2,  we  derived  the  evolution  of  a pure   gas  initial
perturbation, while in  section 3, we derived  the evolution of a pure
dark matter one,  using  for both case  the same  initial profile.  We
conclude that these   two cases give different  results  for  the mass
distribution within the cluster  ($\rho \propto r^{-12/7}$ for gas and
$\rho \propto r^{-2}$ for dark matter). In this  section, the fluid we
consider is a mixture of gas and dark matter particles.  This leads to
a new important parameter for the dynamics of the forming cluster, the
gas  fraction of the  unperturbed  background $\Omega _{B}$ (we  still
assume an Einstein--de Sitter universe   with $h=0.5$, but our  results
are valid for any cosmological parameters).

\subsection{Halo Structure}

For a  dark matter dominated universe (with  $\Omega _{B}  \ll 1$), it
seems  natural to  obtain for  the   gas density profile very  similar
results than for  dark matter. To test this  hypothesis,  we run three
simulations with the same initial parameters

\begin{equation}
M = 10^{16}~M_{\odot}~~~z_c = 2~~~T_0 = 0.014~\hbox{K}
\end{equation}

\noindent 
allowing the baryon density parameter to take three different values

\begin{equation}
\Omega_B = 0.01~~~~\Omega_B = 0.9~~~~\Omega_B = 0.99
\end{equation}

\noindent 
We choose a low initial temperature, which, as we know from section 2,
leads to a very small core ($r_{core} \simeq 10^{-3}$ kpc). This is of
course very far from the observed values in large X-ray clusters where
$r_{core} \simeq   100$  kpc.  We  are  interested  here in  the  halo
structure, where a power law is observed for both  dark matter and gas
density profiles.  The core problem will be addressed below.

\begin{figure}[httb]
%\hspace{-1cm}
\hbox{\psfig{file=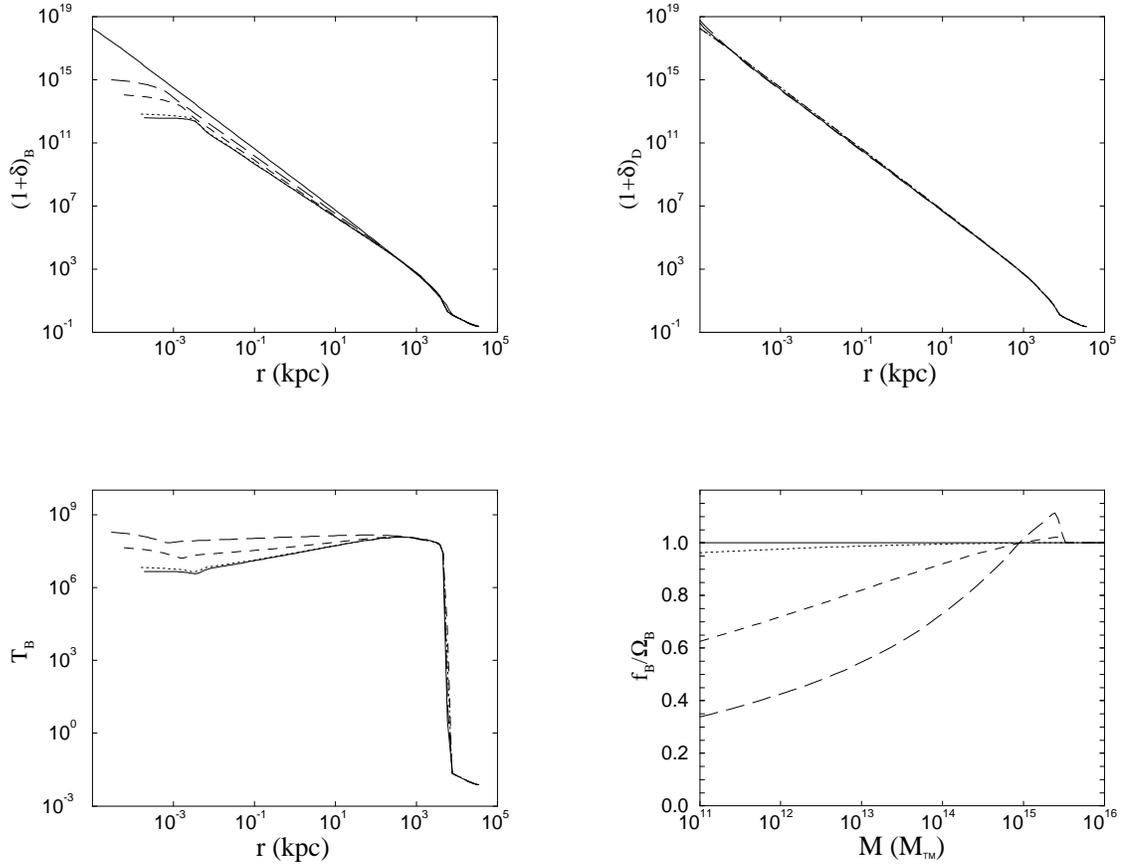,height=13cm,angle=270}}
\caption{Dynamical structure   of a spherical cluster   at z =  0 with
various $\Omega_B$.  (a) Gas over-density  profiles for $\Omega_B$ = 1
(solid line),    0.99  (dotted line),   0.9 (dashed    line)  and 0.01
(long-dashed line).  We  also  plot for   comparison the dark   matter
over-density profile obtained in the $\Omega_B$ =  0 case (upper solid
line). (b)   Dark matter over-density profile  for  the same values of
$\Omega_B$. Note that all  profiles are similar to  the $\Omega_B$ = 0
case. (c) Gas temperature for $\Omega_B$ = 1, 0.99, 0.9 and 0.01.  (d)
Gas mass fraction versus total mass for  $\Omega_B$ = 1, 0.99, 0.9 and
0.01.}
\label{figomegaB}
\end{figure}

We  plot  in   Figure   (\ref{figomegaB}) the  gas   and   dark matter
overdensity profiles obtained  at z = 0 for  various $\Omega_B$.  For
sake of comparison, we   also plot the  gas  density profile for   the
$\Omega_B$   = 1 case,   and the dark  matter density  profile for the
$\Omega_B$  = 0 case.   Note that, even for low   values of the baryon
density parameter, the gas  distribution   does not follow the    dark
matter distribution.  The gas density profile is  steeper than for the
pure gas case, but does not fit the dark matter  profile.  This is due
to the presence of a core that prevents gas to fall deeper in the dark
matter   potential well, and to  the  shock relations  followed by gas
(three  internal degrees of freedom)  which differ from those followed
by dark matter  (one internal degree of freedom).   Note also that the
dark matter density profile is insensitive to the value of $\Omega_B$.
The main effect of dark matter on the gas  distribution is to compress
the   core and the halo  after  the collapse  epoch.   This leads to a
strong heating of  the core due  to  adiabatic compression  only.  The
resulting temperature is much higher than for the pure  gas case.  The
slope in  the temperature profile  changes also slightly, leading to a
much more  isothermal  profile.  For  $\Omega_B   \simeq 1$,  the  gas
distribution  recovers the same characteristics  than for the pure gas
case.

One interesting quantity is the  gas fraction,  defined as $f_{gas}  =
M_{gas}(\le r)/M_{tot}(\le r)$, the ratio of the  enclosed gas mass to
the total gravitational  mass.  Because the slope  of the dark  matter
density profile is steeper ($\rho_{dark} \propto r^{-2}$) than for gas
($\rho_{gas}  \propto  r^{-12/7}$), the    gas fraction increase  with
radius.  In the  core, $f_{gas}$  has  a very low value,  because dark
matter dominates.  In the halo,  for a dark matter dominated Universe,
$f_{gas}$   increases  continuously  with  radius as   $r^{2/7}$, with
$f_{gas} \simeq   \Omega_B$ in  the  most outer  regions.   We plot in
Figure  (\ref{figomegaB})  $f_{gas}$ versus the total   mass as in DJF
(see their fig.  [5]).  When we compare our  profile with the observed
ones in the most massive clusters, we find an excellent agreement.

\subsection{The Core Problem}

We saw in section 2 that, for very  general initial conditions, a core
forms with very high density, and we just saw in  this section that if
dark matter is  present,  the central density   is  even higher.   The
observed core density  in massive X--ray  clusters is not  higher than
$10^{-1}\hbox{  cm}^{-3}$, with  core radii   of the  order  of a  few
hundred kpc.  This apparent  discrepancy is due  to  the fact  that in
adiabatic models  the initial entropy   of the primordial  gas is very
low, leading to   very  small cores.   We saw   that within  adiabatic
models, it  was impossible to  generate entropy inside  the core, even
within a dark matter  dominated scenario.  When  cooling is taken into
account, the situation  is worst:  the  resulting decrease  in entropy
naturally leads  to the formation  of  an ultra high  density  central
core,  in   about  a   free--fall time.   It  seems   clear  that  this
uncomfortable situation is unavoidable, if  one restrict to  dynamical
considerations only. On the other hand, it is possible to imagine many
entropy  generating processes, such  as   supernovae driven winds,  UV
background radiation field, ambipolar diffusion, electronic conduction
and so on.  It is far beyond the scope of this paper to treat all this
processes.  To see an interesting review of this question, see Sarazin
(1990).  We simplify  greatly  the  problem  by considering a   simple
model, that we describe now.

We consider  an homogeneous heating  process, such as  a background UV
radiation field  generated  by primordial quasars  or  stars. We model
this process by turning on  gas heating at say z  = 5, and by imposing
the following conditions.  If  $T < 10^4$ in a  cell, we reset the gas
temperature by imposing $T = 10^4$. On the contrary,  if $T \ge 10^4$,
the  gas   temperature in  the  cell   remains  unchanged.   This is a
continuous entropy generating process, which has also physical support
from   the processes we    just  enumerated.  In  this framework,  the
background temperature  is a free  parameter,  and we consider  here 3
different values

\begin{equation}
T_{bg} = 10^4~\hbox{K}~~~T_{bg} = 10^5~\hbox{K}~~~T_{bg} = 10^6~\hbox{K}
\end{equation}

\noindent 
The last value is much higher  than the permitted temperature obtained
by  quasars   spectra,  but it  gives   us a  limit  for  high entropy
generating processes,  such as supernovae driven  winds.   We take the
following parameters for these three runs

\begin{equation}
M = 10^{16}~M_{\odot}~~~z_c = 2~~~\Omega_B = 0.1
\end{equation}

\noindent 
This  method enables  us  to use the  analytical  formulae  derived in
section 2   for    the  pure  gas   collapse, using     $T_0 =  T_{bg}
(1+z_c)^{-2}$.

\begin{figure}[httb]
%\hspace{-1cm}
\hbox{\psfig{file=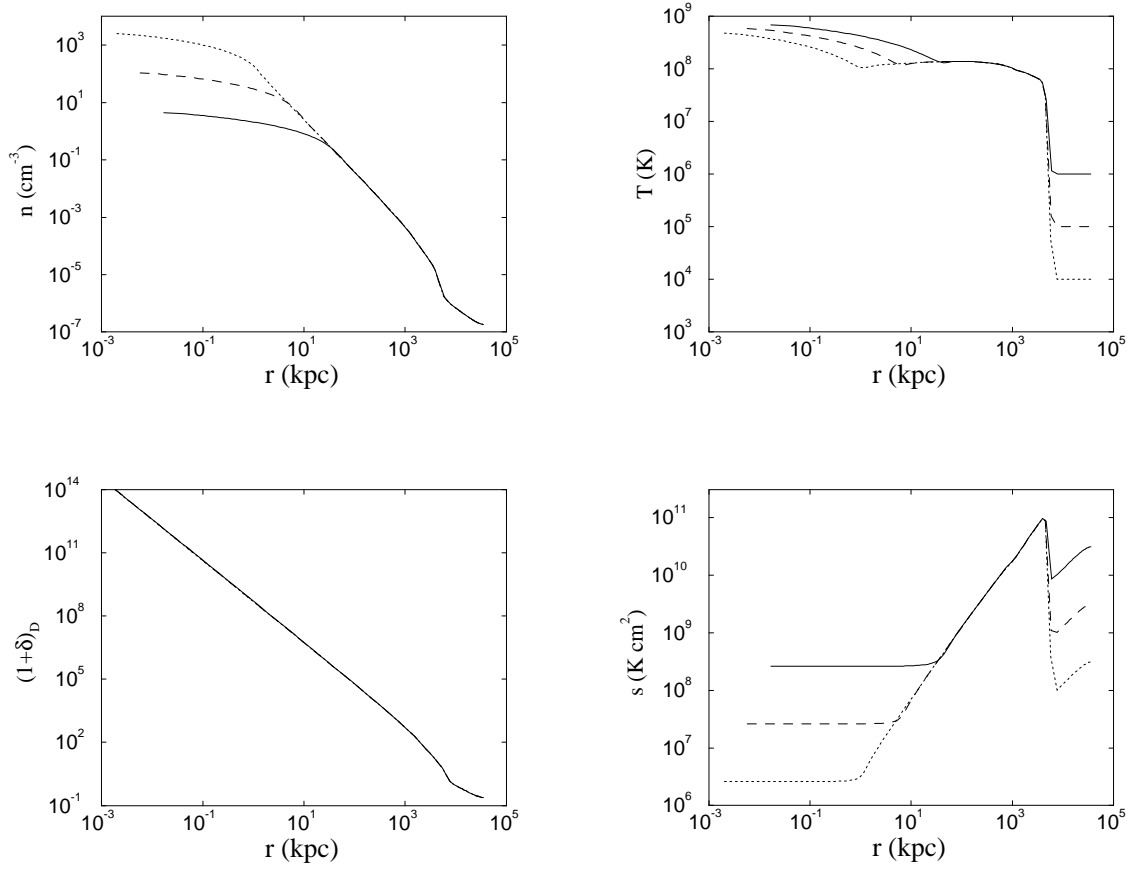,height=13cm,angle=270}}
\caption{Dynamical structure of a spherical cluster ($M = 10^{16}
~M_{\odot}$ and $z_c = 2$) obtained at z = 0 for various 
initial entropy levels. The parameters are (see text) $T_{bg} = 10^4$ K 
(dotted line), $10^5$ K (dashed line) and $10^6$ K (solid line).}
\label{coreheating}
\end{figure}

We plot  in   Figure (\ref{coreheating})  the  resulting  structure we
obtained at z = 0.  Note that the core density is  much lower than for
the other cases.  The core entropy has been greatly enhanced, leading,
as expected, to a more extended core.   Outside the core, the profiles
remain unchanged  by the   heating  process  relative to   the  purely
adiabatic case.  Note also that we predict with our analytical model a
decrease  of the central density by  a factor of  16 between the three
runs, which is exactly what we recover in Figure (\ref{coreheating}).
The dark  matter overdensity profile  is still insensitive to the gas
distribution, as noticed in the last section.

Note that even for the higher temperature  run, the core size is still
unrealistic.   This  is partly  due  to  the  large mass and  collapse
redshift we  use in our simulations.   We believe however that in real
clusters the initial entropy level should be as high as $T_{bg} \simeq
10^{7}$ K, in order to obtain large enough core radii.  An interesting
study,  that we  postpone to  a  next paper, would   be to deduce from
observed X--ray  clusters the required entropy  level  to explain their
core sizes and densities.  This could give important clues to seek for
efficient heating processes in galaxy clusters.

\section{CONCLUSIONS}

The  main goal of this  paper is the study at  very high resolution of
the  collapse of   cosmological structures. This   requirement implies
necessarily   a 1D approach.    Pancakes formation has been thoroughly
studied over the past decade (Zel'dovich 1970; Bond {\it et al.} 1984;
Shapiro \&  Struck--Marcell   1985;  Anninos \&  Norman    1994), while
spherical collapse remained unexplored at an equivalent resolution (in
our case, the central resolution is of the order of 0.1 pc).

In the adiabatic case examined in this  first paper, the main features
of the virialized structures result from the dissipation of supersonic
infall  motions. One therefore should take  care of the method used to
handle  shock waves,  especially  in  spherical  geometry. Within  the
artificial viscosity method, we derive the  form of the viscous forces
one  should use to  get a correct  description of shocks, both in time
and  space. This  general prescription  can   be easily adapted to  3D
hydrodynamics.

This provides  a    clear  account of  the  formation   of  core--halo
structures in the  pure gas case.   During the  collapse, the flow  is
characterized by  a subsonic inner region (the  core), surrounded by a
large envelop where strong  shock dissipation occurs  (the halo).  The
core  and  the halo   are merely the    remnants of the  adiabatic and
dissipative regions of the flow.  The size of  this isentropic core is
therefore  mainly determined by the initial  entropy of the primordial
gas. We obtain for a wide range of initial conditions very small cores
with very high density, definitely ruled out by X--ray observations.

These features  are not qualitatively affected  by the presence of the
dark matter component, but quantitatively the situation is even worse.
The  main effect of dark  matter is to compress  further the gas core,
but with subsonic velocities, so that, again, no entropy is generated.

However,   this  conclusion relies on     the treatment  of the  inner
gravitational field dominated by dark matter. Due to the divergence of
the potential  at the center,  methods using particles are excluded in
practice  by the current computers  limitations.   For that reason, we
propose a fluid approach describing the dynamics  of dark matter. This
method   successfully  reproduces   the  analytical  solutions  of the
self--similar spherical collapse.  However, caustics are  smoothed out
within  this  model,  but the  leading   shell crossing appears   as a
discontinuity in the dark matter  flow at the  correct location.  This
model   provides   moreover a  new insight    in dark matter spherical
dynamics, treated here as a $\gamma = 3$ monodimensional fluid.

We  specifically studied  initial  conditions with  a single spherical
Fourier mode  as initial density  perturbation.  The results we obtain
are   then  valid for  a   wide range  of  arbitrary spherical initial
conditions.  We show analytically  that in the pure  gas case the halo
has a density profile $\rho \propto r^{-12/7}$, while in the pure dark
matter case, it  is close  to  $\rho \propto  r^{-2}$.  The  numerical
results are in  perfect  agreement with  this  scale invariance.   The
specific features of the  coupled  case are  the followings.   We show
that, even for very low $\Omega_B$, the gas never strictly follows the
dark matter particles.  This is   due  to their different numbers   of
translational degrees of freedom.   As   a consequence, the  gas  mass
fraction  is an increasing function of  radius.  Gas  is more smoothly
distributed than dark matter, with a density profile in--between those
obtained for pure gas and pure dark matter collapse.  Adiabatic models
are  then an natural  explanation of  the  observed {\it antibias}  in
X--ray clusters (DJF).   In the   most  outer regions, the gas    mass
fraction reaches the unperturbed value $\Omega_B$.
 
These conclusions can be naturally affected by  the actual geometry of
a  fully 3D gravitational  dynamics,  in which  the collapse occurs in
sheet--like structures, in    filaments and  also  in  quasi--spherical
knots.  This could lead  to different density  profiles, mainly in the
inner  part where rotational  effects   are expected to be  important.
Recent 3D studies show indeed that the mass profile in clusters may be
smoother than  the singular   isothermal sphere  in the  central  part
($\rho \propto r^{-1}$; see Navarro et al.  1995;  Cole \& Lacey 1996;
Tormen,  Bouchet \& White  1996).    However, even  in  3D,  the dense
adiabatic cores are likely to merge subsonically, leaving unsolved the
core problem.  Anninos \& Norman  (1996) investigated the influence of
the actual numerical resolution on a simulated cluster.  They conclude
that the gas central density does not converge, even for their highest
resolution run.  In fact, we show in this paper  that the typical size
of cores obtained for an  adiabatic collapse should be about $10^{-3}$
kpc.  Heating processes are therefore unavoidable to obtain measurable
core radii. Antibiasing has been also investigated in 3D simulations
(Kang et al.  1994; Martel et  al. 1994),  where a strong  segregation
between gas and dark matter has been  detected.  However, Steinmetz \&
White (1996)  have shown that this  effect could be  due to a spurious
heating of the gas,  due to the finite  number of particles used in 3D
simulations.   We  stress here  that the antibias  discussed in this
paper, in  a rather constrained geometry,  has a firm physical origin:
the different dimensionality of gas  ($\gamma = 5/3$) and dark matter
($\gamma =  3$)   phase   spaces.  This  result   was  obtained  using
analytical   or  semi--analytical methods,   free  from any  two--body
relaxation or spurious viscous effects.

We briefly  investigate in this  paper the initial entropy level which
would lead  to  more  realistic  cores properties.   We   find that UV
radiation   fields, accounting only   for a  background temperature of
$10^4$   K    up  to $10^6$  K,    are  unable  to     solve  the core
problem. Consequently, it appears that the  initial entropy level that
could  alleviate this short cut  is very high, and  must result from a
background temperature as high as $10^7$ K.  We therefore analyze in a
companion paper the thermodynamical aspects  of collapse, regulated by
the balance  between   gas cooling and new    energy sources like  the
feedback from star formation.

We  thank  an   anonymous referee for   his   careful reading  of  the
manuscript  and his constructive  comments   which have improved   the
quality of our work.

%\end{references}

%\onecolumn
\pagebreak

%\pagebreak

%\pagebreak

%\pagebreak

%\pagebreak


\begin{thebibliography}{}

\bibitem[Anninos  \& Norman  1994]{Anninos94}Anninos, W.Y., \& Norman,
  M.L., 1994, ApJ, 429, 434

\bibitem[Anninos  \& Norman  1996]{Anninos96}Anninos, P.A., \& Norman,
  M.L., 1996, ApJ, 459, 12
   
\bibitem[Bertschinger 1985]{Bertschinger85} Bertschinger, E.,  1985,
  ApJS, 58, 39
  
\bibitem[Blottiau, Chi\`eze \& Bouquet 1988]{Chieze2} Blottiau, P.,
  Chi\`eze, J.-P. \& Bouquet, S., 1988, A\&A, 207, 24
  
\bibitem[de Boisanger, Chi\`eze \& Meltz 1992]{Chieze3} de Boisanger, C.,
  Chi\`eze, J.-P. \& Meltz, B., 1992, ApJ, 401, 182
  
\bibitem[Bond  et  al. 1984]{Bond84}Bond,  J.R., Centrella, J. Szalay,
  A.S., \& Wilson, J.R., 1984, MNRAS, 210, 515

\bibitem[Briel {\it et al.} 1992]{Briel92} Briel, U.G., Henry, J.P. \&
  Bohringer, H, 1992, A\&A, 259, L31.

\bibitem{} Cioffi,   McKee \& Bertschinger,  1988,   ApJ,  {\bf  334}, 252

\bibitem[Chi\`eze,  Teyssier \&  Alimi 1997]{Paper2} Chi\`eze, J.-P.,
  Teyssier, R., \& Alimi, J.-M., 1997, in preparation.
  
\bibitem[Cole   \& Lacey 1996]{Cole96} Cole,   S.,  Lacey, C.G., 1996,
  MNRAS, in press.

\bibitem[David, Jones \& Forman 1995]{David95} David, L.P., Jones, C.,
\& Forman, W., 1995, ApJ, 445, 578 (DJF).
  
\bibitem[Fillmore  \&     Goldreich    1984]{FG}Fillmore, J.A.,   \&
  Goldreich, P., 1984, ApJ, 281, 1 (FG)
  
\bibitem[Flores \& Primack 1996]{Flores96} Flores, R.A. \& Primack, J.R. 1996,
  ApJ, 457, L5.

\bibitem[]{}Hausman, M.A., Olson, D.W., \& Roth, B.D., 1983, ApJ, 270, 351.

\bibitem[]{}Hoffman, G.L., Olson, D.W. \& Salpeter, E.E., 1980, ApJ, 242, 861.

\bibitem[]{}Hoffman, G.L., Salpeter, E.E., \& Wasserman, I., 1983, ApJ, 268, 
  527.

\bibitem[Hughes 1989]{Hughes89}Hughes, J.P., 1989, ApJ, 337,21. 

\bibitem[]{}Kang, H., Cen, R., Ostriker, J.P., \& Ryu, D., 1994, ApJ, 428, 1. 

%\bibitem[Kofman \& Shandarin   1988]{KS88}Kofman, L.A., \&  Shandarin,
%  S.F., 1988, Nature, 334, 129
  
\bibitem[Landau  \& Lifshitz  1982]{Landau}Landau,  L.D. \&  Lifshitz,
  E.M.,   1982 ``Course  of   Theoretical  Physics,   Volume  6: Fluid
  Mechanics'' (Pergamon Press)

\bibitem[]{} Martel, H., Shapiro, P.R., Valinia, A., \& Villumsen, J.V., 
  1994, in Dark Matter, eds S.S. Holt \& C.L. Bennett, AIP Conference 
  Proceeding 336, p.441.
    
\bibitem[]{} Martel, H., \& Wasserman, I., 1990, ApJ, 348, 1.

\bibitem[Mellier {\it et al.} 1994]{Mellier94} Mellier, Y., Dantel-Fort, M.,
  Fort, B. \& Bonnet, H. 1994, A\&A, 289, L15. 

\bibitem[Moutarde {\it et  al.} 1995]{Moutarde95}Moutarde,  F., Alimi,
  J.-M., Bouchet \& F.R., Pellat, 1995, ApJ, 441, 10
    
\bibitem[Navarro \& White 1993]{Navarro93} Navarro, J.F., White, S.D.M., 1993,
  MNRAS, 265, 271

\bibitem[Navarro,  Frenk  \&  White  1996]{Navarro96} Navarro,   J.F.,
  Frenk, C.S., White, S.D.M., 1995, MNRAS, 275, 720.

\bibitem[Pearce, Thomas, \& Couchman 1994]{Pearce94} Pearce, F., Thomas, P.A.
  \& Couchman, H.M.P., 1994, MNRAS, 268,953.

\bibitem[]{} Peebles, P.J.E., 1982, ApJ, 257, 438.

\bibitem[Rothenflug {\it et al.} 1994]{Chieze1} Rothenflug, R., Magne, N.,
  Chi\`eze, J.-P., Ballet, J., 1994, A\&A, 291, 271
  
\bibitem[Sarazin 1990]{Sarazin90}Sarazin,  C.L., 1990, ``X-ray spectra
  of Clusters of  Galaxies''  (Cambridge University  Press: Cambridge,
  England and New York)
  
\bibitem[Shapiro   \& Struck-Marcell 1985]{Shapiro85}Shapiro, P.R., \&
  Struck-Marcell, C., ApJSS, 57, 205
  
\bibitem[]{}Steinmetz, M., \& White, S.D.M., 1996, submitted to MNRAS, 
  astro--ph/9609021
  
\bibitem[]{}Teyssier,   R., Chi\`eze, J.-P., \& Alimi, J.-M., 1997, 
  ApJ, in press.
  
\bibitem[Thoul and Weinberg 1995]{Thoul95}  Thoul, A.A.,  \& Weinberg,
  D.H., 1995, ApJ, 442, 480


\bibitem[]{} Tormen, G., Bouchet, F.R., \& White, S.D.M., 1996, MNRAS, 
  submitted.

\bibitem[Tscharnuter \& Winkler (1979)]{Tschar79}Tscharnuter, W.-M.,  \&
  Winkler, K.-H., 1979, Comput. Phys. Comm., 18, 171
  
\bibitem[Tyson, Valdes \& Wenk 1990]{Tyson90} Tyson, J.A., Valdes, F. \&
  Wenk, R.A. 1990, ApJ, 349, L1.

\bibitem {} Von Neumann, J., \& Richtmyer, R.D., 1950, J. Appl. Phys.,
  {\bf 21}, 232.

\bibitem[White {\it et al.} 1993]{White93}  White, S.D.M., Navarro, J.F., 
  Evrard, A.E., \& Frenk, C.S. 1993, Nature, 366, 429.

\bibitem[Wu \& Hammer 1993]{Wu93} Wu, X.P., \& Hammer, F. 1993, MNRAS,
  262, 187.

\bibitem[Zel'dovich    1970]{Zeldovich70}Zel'dovich,   Ya.   B., 1970,
  A$\&$A, 5, 84

\end{thebibliography}
\end{document}